\documentclass[aps,pra,twocolumn,showkeys,groupedaddress,nofootinbib]{revtex4-1}

\usepackage{natbib}
\usepackage{graphicx}
\usepackage{dcolumn}
\usepackage{amsmath}
\usepackage{bm}
\usepackage{epstopdf}
\usepackage{verbatim}
\usepackage{float}
\usepackage{lmodern}
\graphicspath{{./figs/}}
\usepackage{hyperref}
\usepackage{stackrel}
\usepackage{xcolor}
\hypersetup{
  colorlinks   = true, 
  urlcolor     = blue, 
  linkcolor    = blue, 
  citecolor   = blue 
}
\usepackage[all]{hypcap}
\definecolor{orcidlogocol}{HTML}{A6CE39}

\begin{document}


\title{Explosion Dynamics of Methane Clusters Irradiated by 38 nm XUV Laser Pulses}

\author{A. Helal  \href{https://orcid.org/0000-0003-3354-6352}{\includegraphics[scale=.6]{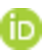}}} 
\email[Corresponding author: ]{ahelal@utexas.edu}
\altaffiliation[Current affiliation: ]{Center for Nonlinear Dynamics (CNLD), The University of Texas at Austin}

\author{S. Bruce}
\author{H. Quevedo}
\author{J. Keto}
\author{T. Ditmire}
\affiliation{Center for High Energy Density Science, The University of Texas at Austin, Austin,TX 78712 USA}
\date{Feb. 2020}

\begin{abstract}
We have studied the explosion dynamics of methane clusters irradiated by intense, femtosecond, 38 nm (32.6 eV) XUV laser pulses. The ion time-of-flight spectrum measured with a Wiley-McLaren-type time-of-flight spectrometer reveals undissociated molecular $\textrm{CH}_4^+$ ions, fragments which are missing hydrogen atoms due to the breakage of one or more C-H bonds $(\textrm{CH}_3^+, \textrm{CH}_2^+ \ \textrm{and}\ \textrm{CH}^+)$ and the recombination product $\textrm{CH}_5^+$. Also visible on the time-of-flight traces are atomic and molecular hydrogen ions $(\textrm{H}^+ \textrm{and}\ \textrm{H}_2^+)$, carbon ions, and larger hydrocarbons such as $\textrm{C}_2 \textrm{H}_2^+$ and $\textrm{C}_2\textrm{H}_3^+$. No doubly-charged parent ions $(\textrm{CH}_4^{2+})$ were detected. The time-of-flight results show that total and relative ion yields depend strongly on cluster size. The absolute yields of $\textrm{CH}^+_5$ and $\textrm{H}^+$ scale linearly with the yields of the other generated fragments up to a cluster size of $\langle\textrm{N}\rangle=70,000 \ \textrm{molecules}$, then begin to decrease, whereas the yields of the $\textrm{CH}_n^+(n=1-4) $ fragments plateau at this cluster size. The behavior of $\textrm{H}^+$ may be understood through the electron recombination rate, which depends on the electron temperature and the cluster average charge. Moreover, the $\textrm{CH}_5^+$ behavior is explained by the depletion of both $\textrm{CH}_4^+$ and $\textrm{H}^+$ via electron-ion recombination in the expanding nanoplasma. 
\end{abstract}
\maketitle

\section{Introduction}

The first observation of high-order harmonic generation (HHG) using tabletop lasers, performed by McPherson et al., opened the door to the widespread experimental study of the interactions between matter and intense electromagnetic radiation at ultraviolet frequencies \cite{Mchp87}. High harmonic radiation can be generated by focusing an ultra-intense femtosecond laser into a rare gas medium, thereby providing an ultrafast source of intense radiation in the extreme ultraviolet (XUV) regime. This high-frequency source exhibits unique characteristics, such as pulse durations in the range of tens of attoseconds \cite{Kapteyn97,Spielmann97,Salieres99}, while maintaining spatial and temporal coherence \cite{Salieres99,Spielmann08,Deroff2000}. These intense XUV pulses interact with matter differently than visible or infrared light due to their higher photon energies. Van der Waals-bound rare gas clusters are a commonly used model target for studies of the interaction of high-intensity laser pulses with nano-scale matter. Their high density and small size make them a unique target for high-intensity laser experiments - exhibiting properties of both gases and solids.
From intense near-infrared (IR) experiments, it is known that, depending on size and electron density, clusters may explode primarily by Coulombic or hydrodynamic forces \cite{Ditmire1997,Ditmire2011}. These two limits exhibit very different cluster explosion times and signatures. The ionization process leading to cluster explosion is strongly wavelength-dependent from IR through XUV \cite{Hoff2011,Bostedt2008} to the X-ray regime; because the kinetic energy of the released electrons determines the charge distribution within the cluster and therefore the explosion dynamics.

When clusters are irradiated with XUV laser pulses, the ponderomotive energy imparted to the electrons is low, but individual photon energies easily exceed the ionization potential of the cluster atoms.  Thus, in this regime, single photon ionization is the dominant ionization mechanism. As electrons are ejected from the cluster, the growing electric potential well results in later-ejected electrons with less kinetic energy than initially-released photoelectrons. This leaves a telltale sign of the  direct multi-step ionization in the photoelectron spectrum \cite{Bostedt2008}. Eventually, the Coulomb field induced by the cluster charge becomes large enough to prevent electrons from escaping the cluster, creating a quasineutral nanoplasma \cite{Arbeiter2010,thomas2009shell}.

Previous experimental investigations into intense, high-frequency laser-cluster interactions have relied on free-electron lasers as the photon source. Several experiments at the free-electron lasers FLASH (DESY facility) \cite{Iwan2012,Timneanu2013} and LCLS (SLAC facility) \cite{Thomas2012} used different intense soft X-ray energies (92 eV at FLASH, and 850 eV at LCLS) to study the explosion dynamics of $\textrm{CH}_4$ and $\textrm{CD}_4$ as well as xenon clusters. The results of the FLASH experiment showed that the explosion dynamics depend on the cluster size, and indicated a transition from Coulomb to a hydrodynamic explosion as the cluster size increased, while the SLAC experiment demonstrated the formation of a xenon nanoplasma that exploded hydrodynamically. A growing number of studies have also focused on theoretical as well as numerical investigation of the explosion dynamics of these large clusters irradiated by intense XUV pulses. 

In this article, we report on an experimental study of the explosion dynamics of methane clusters irradiated by high intensity 38 nm wavelength (32 eV) XUV pulses generated from a tabletop laser system. We were primarily focused on investigating the dependence of the transition from Coulomb to hydrodynamic explosions on the cluster size. We found that the yield of $\textrm{CH}^+_5$ and $\textrm{H}^+$ is qualitatively different from that of the other fragments detected in the experiment, and that prominent transitions in yield-to-cluster-size ratio occur around a cluster size of $\langle\textrm{N}\rangle=70,000\ \textrm{molecules}$, in a good agreement with previous experiments \cite{Iwan2012,Timneanu2013}.

This paper is organized as follows: In Sec. II we introduce the experimental apparatus. The experimental results are described in Sec. III. In Sec. IV we present the results, analysis, and discussion, and the conclusions are drawn in Sec. V.

\section{Experimental setup}
\begin{figure}
\includegraphics{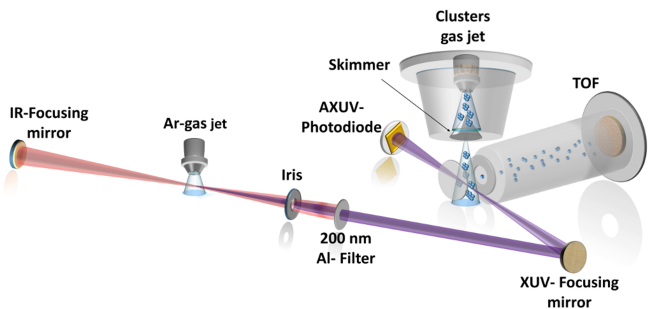}
\caption{\label{figure:1} Diagram (not to scale) of the experimental setup. XUV pulses were generated by focusing 800 nm anular profile laser pulses into a supersonic argon gas jet. The XUV was then separated from the IR using an iris followed by a 200 nm thick Al-filter. The transmitted XUV pulses were then focused into another gas jet (the cluster target) by a spherical Si/Sc mirror, designed to reflect preferentially the $21^{st}$ harmonic.}
\end{figure}

The intense, coherent XUV pulses used in this experiment were generated via HHG by loosely focusing 800 nm ultrafast laser pulses into a supersonic pulsed argon gas jet. The front end of the system was an amplified Ti:sapphire laser, which delivered 600 mJ, 30 fs pulses at a repetition rate of 10 Hz. We applied a mask that, in combination with an inverse mask downstream, separated the generated XUV light from the IR. The scheme removed 17\% of the (previously flat-top profile) beam center, resulting in an annular beam profile with an inner diameter of 12.5 mm and an outer diameter of 30 mm. Fig. \ref{figure:1} shows a schematic of the experimental setup for both the HHG generation and the XUV interaction chamber. The annular beam was brought to a focus by a $f$/200 spherical mirror of focal length 6.0 m, and directed into the supersonic argon gas jet for high-order harmonic generation. We used a pulsed gas source, equipped with a solenoid valve (Parker General Valve, Series 9) and an orifice of 790 $\mu $m. An attached nozzle directed the expanding gas into a slit at its output (0.635 x 6.57 mm), to increase the length of the gas-laser interaction region.

The IR and XUV diverge at different rates with the XUV expanding within the center hole of the IR annulus. At the mask image plane downstream, the inverse mask - an iris - blocked the IR light while allowing the XUV to pass through. Any IR scattered or diffracted in the XUV direction was blocked by a 200 nm thick aluminum filter with 63\% transmission in the XUV for $\lambda < 70$ nm (Luxel Corporation). This two-stage filtering scheme for separating the IR and low harmonics from XUV harmonics resulted in a beam with no detectable IR light on target when measured with a calibrated International Radiation Detectors IRD-AXUV576C photodiode. To ensure that there were no defects in the aluminum filter, a standard microscope slide was inserted regularly into the beam path after the filter, in order to block XUV while transmitting any stray IR. Photodiode measurements of the energy on the target jet indicate no observable IR leakage within the sensitivity of the diode ($<$ 50 pJ/pulse). This corresponded to an IR fluence at the target jet less than 13 $\mu$J/cm$^{-2}$ and an intensity less than $3.9\times10^{8}$ W/cm$^2$, which yielded a measured IR attenuation (aperture + Al filter) of better than $1.3 \times10^{10}$ at the target jet.

After separation from the fundamental frequency, the filtered XUV pulse propagated into the interaction chamber, where it was focused by a $f$/12 Sc/Si multilayer, dielectric-coated, spherical mirror with a focal length of 12 cm. This mirror was designed to reflect only the $21^{st}$ harmonic (38 $\pm$ 5 nm), although aging resulted in some reflection of the neighboring harmonics (discussed in \cite{Hoff2011}). We reflected the incoming beam at about $5^o$ to normal incidence \cite{Uspenskii:98}. We measured the $1/e^{2}$ diameter of the resulting focus to be 7 $\mu $m using the knife-edge technique. We measured pulse energies, for the experiments reported in this article, of (0.6 $\pm$ 0.08 nJ) using a calibrated IRD-AXUV576C photodiode. The diode current was recorded using 50 $\Omega$ termination at the digitizing oscilloscope. The diode charge was determined by numerical integration of the stored time-dependent current. The light pulse energy is then given by the AXUV576C calibrated intensity. Assuming a 7.0 fs pulse duration we estimated an intensity on target of $4.5 \times10^{11}$ W/cm$^2$ \cite{helal2016}.

The targets used in this experiment were van der Waals-bound methane gas clusters, generated via adiabatic expansion of a supersonic gas jet into the vacuum. The jet was injected using a pulsed valve with a conical nozzle with a half-angle of $5^o$, and a throat diameter of 790 $\mu$m. We used the empirical Hagena scaling law \cite{Hagena72} with the modification for larger clusters presented in Arefiev et. al. \cite{Alex2010} to estimate the desired cluster size by varying the backing pressure and nozzle temperature. 
\begin{equation}
\Gamma=\frac{K p d^{0.85}}{T^{2.2875}}     
\label{eq:11} 
\end{equation}
\begin{equation} \phantomsection\label{eq:21}
             <\!\!\textrm{N}\!\!>=
              \begin {cases}
             38.4 (\frac{\Gamma}{1000})^{1.64} & 350< \Gamma<1800 \\ 
             33 (\frac{\Gamma}{1000})^{2.35}  & 1800< \Gamma<10^4 \\ 
             100 (\frac{\Gamma}{1000})^{1.8} &   10^4< \Gamma<10^6  
             \end{cases}
\end{equation}
Where $\Gamma$ is the Hagena parameter, K is the gas sublimation constant (K=2360 for Methane), T is the gas temperature in kelvin and p is the gas backing pressure.

The pulsed nozzle we used was outfitted with a jacket for direct cryogenic cooling. However, for this experiment, the nozzle kept at room temperature and different cluster sizes were obtained by varying the gas backing pressure (0.25 bar $ < p < $ 10 bar) allowing us to reach a wide range of cluster sizes in the range of $4\times10^{2}$ to $3.4\times10^{4}$ molecules per cluster. We used a nozzle opening time of 1 ms to minimize the load on vacuum pumps while reaching near steady-state flow conditions for clustering. The pulsed valve was located inside a small sub-chamber with its dedicated vacuum pumps. As the gas exited the nozzle and expanded, it passed through a skimmer with an orifice 1.2 mm in diameter to create a collimated stream of gas at the interaction region, where the XUV pulses were focused. Particles from the interaction were studied using a Wiley-McLaren type time-of-flight spectrometer \cite{wiley1955} with a drift region of about 0.5 m providing improved mass resolution over previous experiments \cite{Hoff2011}. The length of this drift region was surrounded by a Mu-metal cylinder (wall thickness of 1.575 mm), designed to shield the Earth's magnetic field. The detection system consisted of two microchannel plates in a Chevron configuration for maximum signal to noise ratio. Amplified ion charges were collected by a 50 Ohm impedance-matched anode, where the signals were recorded using a Tektronix TDS5000B - 1 GHz bandwidth, 5 GS/s oscilloscope.

\begin{figure}
\centering
\includegraphics{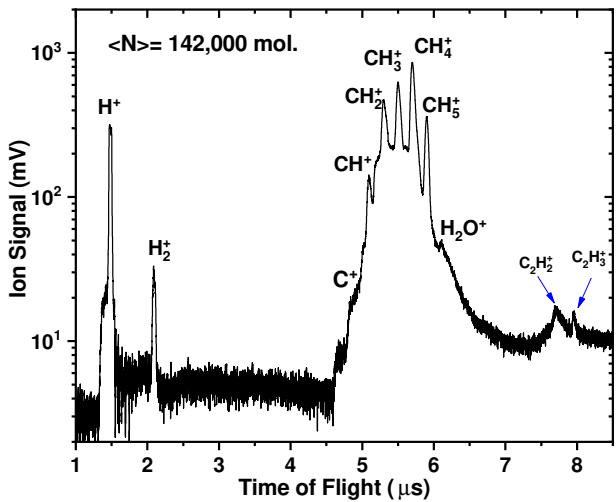}
\caption{\label{figure:2}Methane ion time-of-flight spectrum for a cluster size of $\langle\textrm{N}\rangle=142,000\ \textrm{molecules}$.}
\end{figure}

\begin{figure}
\centering
\includegraphics{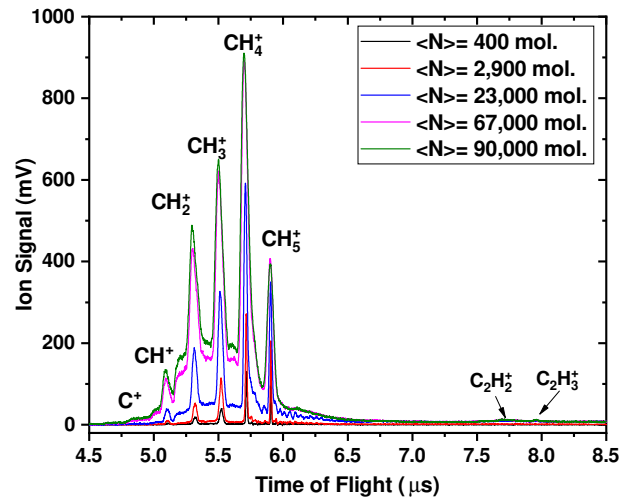}
\caption{\label{figure:3}Methane ion time-of-flight spectra for various cluster sizes.}
\end{figure}

\section{EXPERIMENTAL RESULTS}
The ion time-of-flight spectrum of methane clusters of $\langle\textrm{N}\rangle=142,000\ \textrm{molecules}$, irradiated by 38 nm XUV pulses, is shown in Fig. \ref{figure:2}. The following species can be observed: undissociated molecular ions $\textrm{CH}_4^+$, fragments that are missing hydrogen atoms due to the breakage of one or more C-H bonds $(\textrm{CH}_3^+, \textrm{CH}_2^+ \ \textrm{and}\ \textrm{CH}^+)$, the recombination product $\textrm{CH}_5^+$, atomic and molecular hydrogen ions $(\textrm{H}^+ \textrm{and}\ \textrm{H}_2^+)$, carbon ions, and larger hydrocarbons like $\textrm{C}_2 \textrm{H}_2^+$ \textrm{and} $\textrm{C}_2\textrm{H}_3^+$ molecules. No doubly charged parent ion $(\textrm{CH}_4^{2+})$ or higher charge states of any of the fragments were detected.

\begin{figure}
\centering
\includegraphics{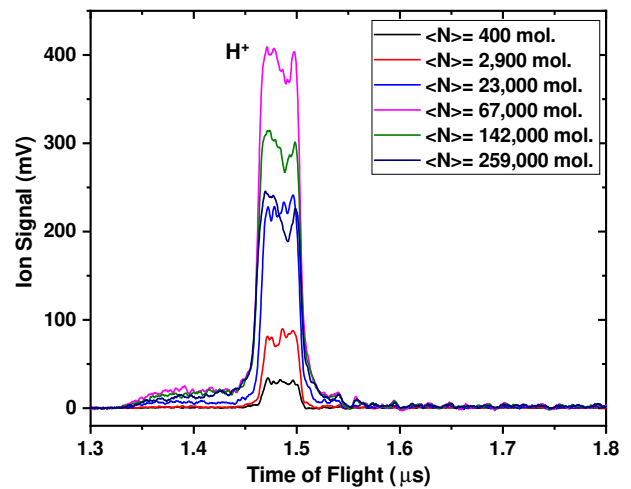}
\caption{\label{figure:4}The yield of the hydrogen peak $H^+$ increases with cluster size up to $\langle\textrm{N}\rangle \simeq70,000 $ molecules, then decreases with increasing cluster size. }
\end{figure}
The change in the ion time-of-flight yields of the methane species as a function of cluster size is shown in Fig. \ref{figure:3}. We observed an increase in the amplitude of the continuum spectra under the methane fragmentation peaks. As found in other gas cluster experiments (Ar, Xe, and $\textrm{N}_2$) \cite{helal2016}, the increase in cluster size leads to a disproportionate increase in the tails of the kinetic energy distributions. In the methane case, the close charge-to-mass ratios of the fragmentation products cause their energy distributions to overlap. A closer look at the ion spectra reveals that the generation of the larger hydrocarbons $(\textrm{C}_2 \textrm{H}_2^+\ \textrm{and}\ \textrm{C}_2 \textrm{H}_3^+)$ occurs only after the cluster size reaches $\langle\textrm{N}\rangle=4,000\ \textrm{molecules}$. Moreover, we noticed that the yield of the $\textrm{H}^+$ ions (Fig. \ref{figure:4}) depends on the cluster size, such that the yield increases with the cluster size up to $\langle\textrm{N}\rangle \simeq70,000 $ molecules, then begins to decrease with further increase in the cluster size.

To understand these observations we present in Fig. \ref{figure:5} the measured ion yield, normalized to the species maximum after removing the continuum spectrum, as a function of the cluster size. We note that the normalized yield of all the fragments and the undissociated $\textrm{CH}_4^+$ shows an increase with cluster size up to $\langle\textrm{N}\rangle\approx 70,000\ \textrm{molecules}$, beyond which it is constant. However, the normalized yield of the recombined $\textrm{CH}_5^+$ and the hydrogen peak $\textrm{H}^+$ do not level off but decrease together at the same rate. This change in yield for $\textrm{CH}_5^+$ was observed before by B. Iwan et al. \cite{Iwan2012} and N. T\^{i}mneanua et al. \cite{Timneanu2013} using the FLASH free-electron laser which produces 15-fs pulses at a wavelength of 13.5 nm (photon energy = 92 eV) and a focused beam fluence of 13.7 J/cm$^2$. The maximum yield in their case occurred for a cluster size of 20,000 molecules. In those experiments, they also found a monotonic increase in the yield of atomic fragments $\textrm{H}^+$ and $\textrm{C}^+$ with increasing cluster size, unlike in this experiment at 32 eV photon energy where $\textrm{H}^+$ behaves similarly to $\textrm{CH}_5^+$. Their trend was linear with the stagnation pressure used to generate the clusters, consistent with the increase in the overall molecule density. Moreover, at the largest cluster size, the $\textrm{H}^+$ fragment accounted for 60\% of the total ions.

To help understand our experimental results, we also analyzed the yield of each measured ion fragment as a fraction of the total ion yield shown in Fig. \ref{figure:6}. The relative yields of $\textrm{CH}^+$, $\textrm{CH}_2^+$, $\textrm{CH}_3^+$ and $\textrm{CH}_4^+$ show a small increase with increasing cluster size up to $\simeq$ 70,000 molecules, where they leveled off at a different constant value for each fragment. Both $\textrm{CH}_5^+$ and $\textrm{H}^+$ ions showed a monotonic decrease in the relative yield with increasing cluster size (especially for $\textrm{H}^+$). Only a small concentration of $\textrm{H}_2^+$ was also observed, which also decreased slightly as the cluster size increased.

\section{Discussion and analysis}
The relative abundances of the methane ion fragments are controlled by dissociative photoionization, dissociative ionization by electron impact, and electron-ion recombination. In this section, we discuss each of these mechanisms separately to understand the change in the behavior of $\textrm{CH}^+_5$ and $\textrm{H}^+$ after a cluster size of $\simeq$ 70,000 molecules, relative to the rest of the generated fragments shown in Fig. \ref{figure:6}. 
\subsection{Dissociative photoionization}
 For smaller cluster sizes, the populations are explained to first order by dissociative photoionization cross sections, measured as a function of photon energy by Samson et al.\cite{samson1989ionization}:
 \begin{subequations}
\begin{eqnarray} 
&CH_4 + h \upsilon \rightarrow{\  } CH_x^+ + (4-x)H + e \label{subeqn-3:1}\\
&CH_4 + h \upsilon \rightarrow{\  } H^+ + CH_3 + e \label{subeqn-3:2}
\end{eqnarray}
\end{subequations}
where $x$ ranges from 0 to 4. Their measured branching fractions for the ions $\textrm{CH}_4^+$, $\textrm{CH}_3^+$, $\textrm{CH}_2^+$, $\textrm{CH}^+$, $\textrm{C}^+$, $\textrm{H}^+$, and $\textrm{H}_2^+$ at 32 eV photon energy were 0.40, 0.43, 8.0$\times10^{-2}$, 3.8$\times10^{-2}$, 7.4$\times10^{-3}$, 4.6$\times10^{-2}$, 3.8$\times10^{-3}$ respectively. Similar branching fractions were observed at a FLASH experiment with $h\upsilon$= 92 eV, for all species other than $\textrm{H}^+$. The Fluence during the FLASH experiment was 4400 times that of our experiment, which makes the similar branching fractions surprising. However, the much smaller photoionization cross-section at 92 eV compared to 32 eV results in an ionization rate only 60 times larger in the FLASH experiments than in ours. This difference partially explains their greater production of $\textrm{H}^+$

\begin{figure}
\includegraphics{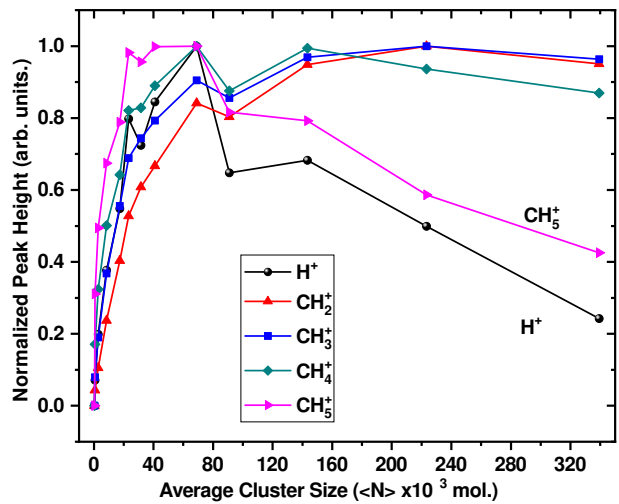}
\caption{\label{figure:5}Measured ion yield normalized to the maximum of each fragments and hydrogen peaks. A high correlation between the yield of $\textrm{H}^+$ \textrm{and} $\textrm{CH}_5^+$ is observed.}
\end{figure}

\begin{figure}
\includegraphics{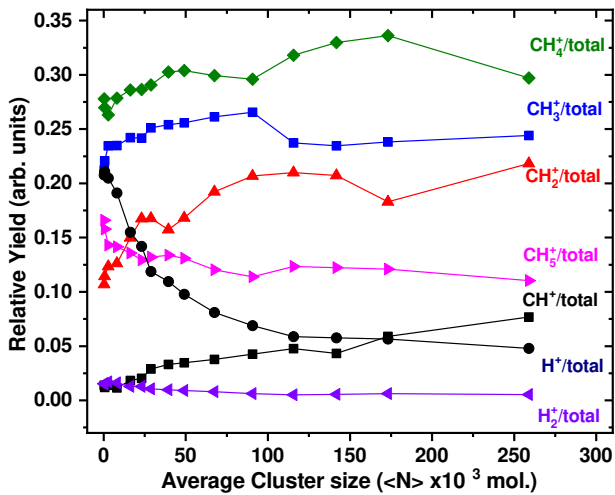}
\caption{\label{figure:6}The relative ion yield of each methane fragments to the total yield as a function of the cluster size. }
\end{figure}

In the previous photoionization experiments by Samson et al. the abundance of the $\textrm{CH}_3^+$ fragment was $\simeq$20\% greater than $\textrm{CH}_4^+$, unlike the experiments both here and at FLASH  \cite{Iwan2012,Timneanu2013}. This difference is likely explained by measurements of Latimer et al., who observed that the $\textrm{CH}_4^+$/$\textrm{CH}_3^+$ ratio increased as the temperature of the sample was decreased to 100 K  \cite{latimer1999dissociative}. The vibrational temperature of $\textrm{CH}_4^+$ in clusters formed in supersonic jets at similar conditions as our experiment is estimated to be closer to 40 K \cite{bartell1986diffraction}. This temperature dependence may explain the 20\% increase in the $\textrm{CH}_4^+$/$\textrm{CH}_3^+$ ratio observed in Fig.\ref{figure:6} for a cluster size larger than $10^5$ molecules. The increase in $\textrm{CH}_4^+$ yield comes primarily at the expense of $\textrm{CH}_3^+$.

\subsection{Dissociative ionization by electron impact}
Dissociative ionization by electron impact also contributes to the ion yield fractions \cite{straub1997absolute}, especially as the cluster size increases. For the 20 eV photoelectrons expected and measured in our experiments the mean free path for elastic scattering is 0.73 nm and for ionization is 5.1 nm\footnote{For momentum transfer cross sections see \cite{davies1989measurements}.}. This range corresponds to a mean size of $\sim$11,000 molecules. These electrons will escape smaller clusters, positively charging them until the cluster potential exceeds the photoelectron energy. The ions and electrons in the center of the cluster will form a neutrally charged plasma. Our photoelectron energy is only above threshold for $\textrm{CH}_4^+$, $\textrm{CH}_3^+$, and $\textrm{CH}_2^+$ ions. Threshold for electron impact dissociative ionization to form $\textrm{H}^+$ is 22.4 eV \cite{straub1997absolute}. In experiments reported here, free protons are produced by photodissociation during the 7 fs laser pulse, and not by electron impact either during the laser pulse or at later times as the cluster explodes.

The photoelectron energy is the major expected difference compared to the FLASH experiments. There, the photoelectron energy of 80 eV, much greater than 20 eV in our case, leads to a mean free path for electron impact ionization of 0.28 nm, with a much greater ionization rate than ours, particularly for H$^+$.  This partly explains the smaller number of $\textrm{H}^+$ ions produced in the current experiments compared with FLASH experiments, where the ion yield fraction for $\textrm{H}^+$ was near 60\% for the largest clusters.

\subsection{Electron recombination}
The electron recombination rate also plays an important role in the difference between experiments at different photon energies. The main recombination mechanism of $\textrm{H}^+$ is the three-body recombination process described by
\begin{equation}\phantomsection \label{eq:4}
H^+ + 2 e \stackrel[]{k_1}{\rightarrow} H + e.
\end{equation}
where the recombination rate, $\textrm{k}_1$ is given by
\begin{equation} \phantomsection\label{eq:5}
k_1=\frac{8.75 \times10^{-27} Z^3}{T_e^{4.5}}   \ cm^6s^{-1} 
\end{equation}
where T$_e$ is the electron temperature in electron volts, and Z is the ionic charge state \cite{hinnov1962electron}. Since the fractional ionization approaches 2\% in our experiments, the electron-ion recombination rate is very fast at the resulting large electron density. The ions observed in the TOF are the sum over the full explosion time  \footnote{In this context, explosion time is defined as the time required by the expanding cluster to reach twice its initial diameter.} for the cluster, and as the cluster size becomes bigger the explosion time increases, and a greater number of ions and electrons recombine. For Z = 1 and T$_e$ $\approx$ 1 eV, which we estimated after adiabatic cooling of the electrons from cluster expansion, the electron density and $\textrm{H}^+$ density will decrease an order of magnitude during a time of 56 fs after the end of the laser pulse. This contributes to the loss in $\textrm{H}^+$ yield observed in Fig. \ref{figure:5} as the cluster size is increased above 70,000 molecules. For the experiments at FLASH, T$_e$ = 4 eV, which reduces the recombination rate by a factor of about 500 \cite{Timneanu2013}. At this higher T$_e$, the recombination time exceeds the explosion time. The difference in the $\textrm{H}^+$ ionization rate and recombination rate at the two photon energies explains the different relationships between $\textrm{H}^+$ concentration and cluster size. For photon energies of 92 eV, the electron impact ionization rate overcomes the slower recombination rate, so that the ion concentration increases with the linearly increasing number of methane molecules in the focal volume. In contrast, our lower photon energy, lower photoelectron energy, larger recombination rates, and longer expansion times decrease the number of surviving $\textrm{H}^+$ with increasing cluster size.

\subsection {The behaviour of $\textrm{CH}_5^+$}
Explaining the variation in population of the larger hydrocarbon ions ($\textrm{CH}_5^+$, $\textrm{C}_2\textrm{H}_2^+$, and $\textrm{C}_2\textrm{H}_3^+$) with cluster size is more difficult. Because of the correlation observed in the behavior of $\textrm{CH}_5^+$ and $\textrm{H}^+$, one is tempted to associate the decrease in $\textrm{CH}_5^+$ ion signal with an increased destruction rate with cluster size, similar to the explanation for $\textrm{H}^+$. The dissociation of $\textrm{CH}_5^+$ by electron collisions is of fundamental importance and is one of the mechanisms that could contribute to the loss of $\textrm{CH}_5^+$ \cite{semaniak1998dissociative,larson1998branching}. Direct dissociative excitation rates are small for electron energies below 10 eV compared to dissociative recombination rates with electrons for energies below 0.2 eV:
\begin{subequations}
\begin{eqnarray} \label{eq:6}
CH_5^++e \stackrel[]{k_2}{\rightarrow}
&CH_4 + H + 8\ eV\\
&CH_3 + H_2 + 7.99\ eV\\
&CH_3 + H +H + 3.51\ eV\\
&CH_2 + H_2+ H + 3.18\ eV\\
&CH + H_2 + H_2 + 3.29\ eV
\end{eqnarray}
\end{subequations}

However, recombination of all of the molecular ions is expected to be dominated by three-body reactions similar to Eq. (\ref{eq:4}) for electron densities greater than 2$\times10^{18} \textrm{cm}^{-3}$ \cite{stabler1963capture}. The time required (21 ps) for the electron density to decay to this concentration exceeds the cluster explosion time unless T$_e$ falls substantially below 1 eV. Recombination is also expected to increase the electron temperature. Qualitatively, over the explosion time, all of the ions are expected to recombine at similar rates. $\textrm{CH}_5^+$ in particular has exactly the same dissociative recombination and three-body rate as $\textrm{CH}_4^+$ \cite{mul1981merged}. Clearly, the difference in the yields of $\textrm{CH}_4^+$ and $\textrm{CH}_5^+$ depends on their production rates and not their recombination rates.

$\textrm{CH}_5^+$ is formed by an association of molecular ions with neutral molecules\cite{herman1990beam,sefcik1974methanium}:
\begin{subequations}
\label{allequations}
\begin{eqnarray}
CH_4^+ + CH_4 \stackrel[]{k_3}{\rightarrow} CH_5^+ + CH_3  \label{subeqn-7:1}\\
H^+ + CH_4 + CH_4 \longrightarrow CH_5^+ + CH_4 \label{subeqn-7:3}\\
CH_4^+ + H_2 \longrightarrow CH_5^+ + H \label{subeqn-7:2}
\end{eqnarray}
\end{subequations}
where the dominant reactions in our experiments are Eqs. (\ref{subeqn-7:1}) and (\ref{subeqn-7:3}) since here the $\textrm{H}_2$ concentration as determined from the $\textrm{H}_2^+$ concentration shown in Figs. \ref{figure:5} and \ref{figure:6} is too small to contribute.

Given the large production rate for $\textrm{H}^+$ in Eq. (\ref{subeqn-3:2}) and the correlation between the $\textrm{CH}_5^+$ and $\textrm{H}^+$ ion yields, it is possible to suggest that the formation channel for $\textrm{CH}_5^+$ is mostly by the association of $\textrm{H}^+$ with methane described by Eq. (\ref{subeqn-7:3}). At solid densities, the three-body rate is expected to be fast and the process could be two-body capture with rotation and vibrational excitation followed by energy loss by phonon production in the solid cluster. The additional loss of both ions by recombination with electrons as the cluster size increases would decrease the production rate of $\textrm{CH}_5^+$. Consequently, formation of $\textrm{CH}_5^+$ would contribute to the loss of both $\textrm{H}^+$ and $\textrm{CH}_4^+$.

To illustrate, we can assume reaction (\ref{subeqn-7:1}) alone contributes to the production of $\textrm{CH}_5^+$, combine this with the loss rate controlled by three-body recombination similar to Eq. (\ref{eq:4}), and predict the concentration of $\textrm{CH}_5^+$, assuming steady-state, to be of the order:
\begin{equation} \phantomsection \label{eq:8}
[CH^+_5]\approx\frac{k_3 [CH^+_4] [CH_4]}{k_1[e]^2}.
\end{equation} 

We would expect an increase in $\textrm{CH}_5^+$ yield because of the increasing number of $\textrm{CH}_4^+$ ions with increasing cluster size, as observed in Fig. \ref{figure:5}. We also observe in Fig. \ref {figure:5} that at a cluster mean size of $\langle\textrm{N}\rangle\simeq$ 70,000 molecules the number of $\textrm{CH}_4^+$ ions saturate, suggesting that their production rate no longer increases with the number of target methane molecules. Above this cluster size the production rate in the numerator of Eq. (\ref{eq:8}) no longer increases to compensate the increasing recombination rate as the cluster size and explosion time increases, and hence the yield decreases. 

\begin{figure} 
  \centering
  \includegraphics{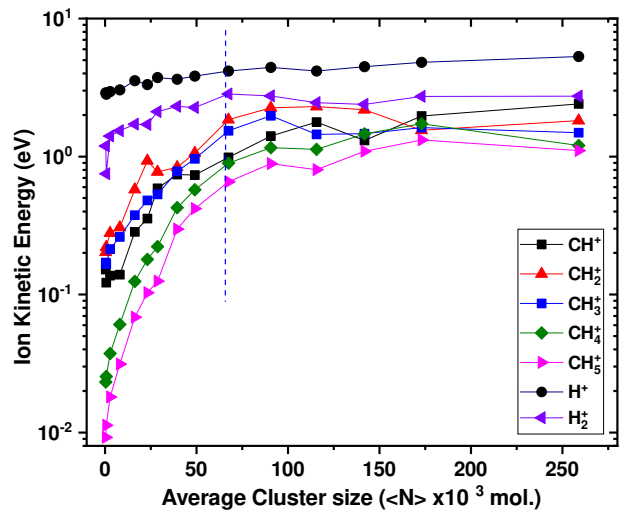}
  \caption{\label{figure:7}The kinetic energy of the generated ions from the explosion of methane clusters, calculated from the time-of-flight peak width.}
\end{figure}

The average kinetic energies of ions originating from the photoionization of methane clusters, derived from the width of the time-of-flight ion peak are shown in Fig. \ref {figure:7}. The kinetic energies of the undissociated $\textrm{CH}_4^+$, the fragments ($\textrm{CH}^+$, $\textrm{CH}_2^+$, and $\textrm{CH}_3^+$), the recombined $\textrm{CH}_5^+$, the molecular $\textrm{H}_2^+$ and $\textrm{H}^+$ all increase as a function of the cluster size up to $\langle\textrm{N}\rangle\simeq 70,000\ \textrm{molecules} $, indicated by the dashed line in the figure. Above this size, the energies remain constant as the cluster size increases, revealing a change in the cluster's explosion dynamic at large cluster size. One of the characteristic differences between a Coulomb explosion and a hydrodynamic expansion is that the latter is characterized by the trapping of electrons inside the cluster, which in turn increases the electron density within the cluster, leading to an increase in the recombination rate for all ions. This change in the explosion dynamics affects both the ion energies and ion chemistry.

\section{Summary}
In summary, we have studied the explosion dynamics of methane clusters irradiated by 38 nm XUV laser pulses. The time-of-flight spectrum shows the undissociated molecular ion $\textrm{CH}_4^+$, fragments without hydrogen atoms due to the breakage of one or more C-H bonds $(\textrm{CH}_3^+, \textrm{CH}_2^+ \ \textrm{and}\ \textrm{CH}^+)$, the recombination product $\textrm{CH}_5^+$, hydrogen ions and molecules $(\textrm{H}^+ \textrm{and}\ \textrm{H}_2^+)$, and carbon ions and larger hydrocarbons such as $\textrm{C}_2 \textrm{H}_2^+$ and $\textrm{C}_2\textrm{H}_3^+$ molecules. No doubly-charged parent ion $(\textrm{CH}_4^{2+})$ was detected. The TOF results show that total and relative ion yields depend strongly on cluster size. $\textrm{CH}^+_5$ and $\textrm{H}^+$ both show a behavior similar to the rest of the generated fragments up to a cluster size of $\langle\textrm{N}\rangle\simeq 70,000\ \textrm{molecules} $, beyond which they exhibit a drop in the yield.  In the other fragments, we observe a plateau in the yield with increasing cluster size, beginning at this same transition point.

A comparison between the data presented in this experiment and that acquired on FLASH with a different photon energy (92 eV) provided much insight into the effect of the photon energy on the cluster dissociation behavior. We were able to explain the behavior of $\textrm{H}^+$ through the electron recombination rate which depends on the electron temperature and the cluster average charge. Moreover, the $\textrm{CH}_5^+$ behavior was successfully determined and explained by the loss of both $\textrm{CH}_4^+$ and $\textrm{H}^+$ by recombination with electrons above a cluster size of $\langle\textrm{N}\rangle\simeq 70,000\ \textrm{molecules} $. This increase in the recombination rate could be explained by a change in the cluster explosion dynamics from Coulomb to hydrodynamics as suggested by the kinetic energy measurement shown in Fig. \ref{figure:7} and the scattering range for photoelectrons produced by 32 eV photons.

\section*{Acknowledgments}
This work was supported by the DOE Office of Basic Energy Sciences and the National Nuclear Security Administration under Cooperative Agreement NO: DE-NA0002008.


\section*{}
\bibliographystyle{apsrev4-1}
\bibliography{Methane-PRA}

\end{document}